# A COMBIMATORIAL ALGORITHM TO GENERATE ALL SPANNING TREES OF A WEIGHTED GRAPH IN ORDER OF INCREASING COST

Barun Biswas#1, Krishnendu Basuli*2, Saptarshi Naskar*2, Saomya Chakraborti*2

Samar Sen Sarma*2

1#West Bengal State University, West Bengal, India

Barunbiswas9u6@gmail.com

*2West Bengal State University, West Bengal, India

Krishnendu.basuli@gmail.com

*2Sarsuna College, West Bengal, India

sapgrin@gmail.com

#2University Of Calcutta, West Bengal, India

itzsoumya@yahoo.com

#University Of Calcutta, West Bengal, India

Sssarma2001@yahoo.com

*Abstract- The most popular algorithms for generation of minimal spanning tree are Kruskal's[2] and Prim's[2] algorithm. Many algorithms have been proposed for generation of all spanning tree. This paper deals with generation of all possible spanning trees in increasing cost of a weighted graph. This approach uses one matrix called Difference Weighted Circuit Matrix (DWCM); it is little bit modification of FCM.*

## I. INTRODUCTION

Finding a minimum spanning tree for a connected weighted graph with no negative weight can be obtained using classical algorithms such as Prims and kruskal. Both of two gives the single minimum spanning tree. But sometimes it needs to generate the second minimum spanning tree, third, fourth and so on. A number of algorithms have been proposed to enumerate all spanning trees of an undirected graph[].Good time and space are the major concern of these algorithms. But there are very few existing algorithms for generating of spanning trees in order of their weights[1,4].This is because the cost of the edges of the graph are not taken into consideration during the generation of the spanning trees. Any of the algorithms[1,4] which generate all spanning trees without weights can be applied to our problem by sorting the trees according to the increasing weight after they have been generated .As the number of the trees are very large this option is excluded for practical purpose. So we propose an algorithm for generation of spanning trees according to their increasing cost in each iteration by using the fundamental circuit matrix concept .And very few spanning trees are stored in practical situation..

## II. BASIC DEFINITION

A. **Graph:**

An undirected graph G=(V,E)consists of a set of objects V={v1,v2,v3....Vn}called vertices and another set E={e1,e2,....}whose elements are called edges, such that each edge $e_k$ is identified with an unordered pair (vi, vj) of vertices.[1,2]

B. **Weighted Graph:**

A weighted graph is a graph G in which each edge e has been assigned a real number w(e) called the weight of e. If H is a sub graph of a weighted graph, the weight w(H) of H is the sum of the weights w(e1) +......+w(ek) where {e1,e2,....ek} is the set of edges of H.

C. **Tree:**

A tree is a sub graph of G that does not contain any circuits. As a result there is exactly one path from each vertex in the tree to each other vertex in the tree.

D. **Spanning tree:**

A spanning tree of a graph g is a tree containing all vertices of G.

E. **Minimum Spanning tree(MST):-**

A minimum spanning tree of an undirected weighted graph G is a spanning tree of which the sum of the edges weights is minimal.

F. **Fundamental Circuit Matrix:**

A sub matrix (of a circuit matrix)in which all rows correspond to a set of fundamental circuit is called a fundamental circuit matrix[2,4].

G. **Branch:**

An edge in a spanning tree of a graph is called the branch.

H. **Cord:**

The edges that are not in the spanning tree of a graph are called the chord. That is the sub graph S' is the collection of Chord of the graph G with respect to S the Spanning tree of the graph.

I. **DWCM:**

The abbreviation is Difference Weighted Circuit Matrix. It is the little bit of modification of the FCM.

A sub matrix in which all rows correspond to a set of fundamental circuits is called a Fundamental circuit matrix. If n is the number of vertices and e is the number of edges in a connected graph, then the matrix is an (e-n-1).(n-1) matrix .Hare the branches weight are present on the column head as branch mark and the chords (e-n-1)are for the row representation. And in the each cell of the matrix is assigned difference weight of the chord and the branches participating for generation of circuit when this chord is joined to the spanning tree presented on the column head.

This structure is used for the proposed algorithm.

$C[ij]=w(c_I)-w(b_j)$ where bj is the branch participating in the

circuit when ci is joined to the present spanning tree presented by the column weighted branch.

=0    Otherwise

Where C[ij] is the value of the cell of the DWCM matrix's $I^{th}$ row and $j^{th}$ column. $w(c_i)$ is the weight of the cord of row 'I' and $w(b_j)$ is the weight of the branch of column 'j'

### III. PREVIOUS WORKS TO DETERMINE MINIMAL SPANNING TREE

There are several greedy algorithms for finding a minimal spanning tree M of a graph. The algorithm Prim and Kruskal are well known.

A. **Kruskal's Algorithm[c/H]:[5][6]**

Kruskals algorithm is one of the optimized way to determine the minimal spanning tree in a connected graph. It always results the optimal solution. The basic steps to determine the minimal spanning tree inn this process is as follows.

Step1:- choose e1 an edge of G ,Such that w(e1) is as small as possible and e1 is not a loop.

Step2:-If edge e1,e2,....ei have been chosen ,then choose an edge ei+1 not already chosen such that

i) The induced sub graph G[{e1,......ei+1}] is acyclic and

ii) ii) w(ei+1) is as small as possible(Subject to Condition (i))

iii) Step3:-If G has n vertices, stop after n-1 edges have been chosen.

iv) Otherwise Repeat Step2.

We know that Kruskal's algorithm generates a minimum cost spanning tree for every connected undirected graph G.

### IV. PROPOSED ALGORITHM FOR GENERATION OF SPANNING TREES IN ORDER OF INCREASING COST

In the following we generate a spanning tree using Prim's or Kruskal's algorithm which gives the Minimum Spanning Tree of the graph G. Then we apply the one of the three available techniques ( elementary tree transformation, Decomposition and Test and Select method) the first one(elementary Tree Transformation Techniques).at a time we replace one of the chord to the any one branch of the spanning tree which makes minimum increment to the total cost of the spanning tree. We basically give importance to the fundamental circuit of the graph. If we follow the techniques of the paper of K. Sorensen, G. Janssens, 2005[5] there need to track the all edges combination on the different spanning trees. So for large graph it become more complicated in practical situations and there needs to generate minimum spanning trees on some bounded condition. So we propose an alternative algorithm to generate all the spanning trees in order of increasing order where very few spanning trees need to save and time complexity under limit in practical.

### A. Algorithm OMST (Ordered Minimal Spanning Trees):

**Input:** E, cost n, m, arr.

// E is the set of edges in G. cost [1: n] [1: n] is the cost of the edge.

// n is the number of vertices and m is the number of edges.

// The spanning trees are computed in order and stored replacing the previous in the array arr[1:n]. The final cost corresponding to the spanning tree is printed.

//The n-1 edges of the minimal spanning trees are termed as **branches** and the rest of the m-n+1 edges of G are termed as **chords**.

**Step 1:** // Difference Weighted Circuit Matrix of the minimal spanning tree.

  i. Insert the branches of the last minimum spanning tree

     stored in arr, at the DWCM column heads.

 ii. Insert the chords as DWCM row heads.

   iii. For i = 1 to n-1, repeat step iv to step vi

   iv. For j = 1 to m-n+1, repeat till step vi

    v. If $i^{th}$ row head edge makes cycle with the edges in

       arr, then do

   vi. Find difference of the cost of the $j^{th}$ DWCM column

       head edge from the $i^{th}$ DWCM row head edge.

**Step 2:** Find minimum element in DWCM>0.

**Step 3:** Add element to the cost of the last minimum tree found.

**Step 4:** Store all column head edges in arr after replacing the edge in column head of the minimum element found with the row head edge of the minimum element position.

**Step 5:** Print the edges along with the cost.

**Step 6:** While all the elements of DWCM are not 0, then do

**Step 7:** For all the positive elements in the column of the minimum value found,

   repeat  Step ii to Step v.

   Add element to the cost of the tree whose DWCM is evaluated.

   Find the edges of the new tree by replacing the

   DWCM row and column head edges.

   Store the cost along with the edges of the new tree evaluated.

   Find minimum among the unused rows of DWCM>0 and Go to Step 7.

**Step 8:** Execute Step 1 to 4.

**Step 9:** If new cost found > minimum of the stored intermediate trees, then

   *do* Step 10 and Step 11

**Step 10**: Execute step 7.

**Step 11:** Execute step 1 to 5.

**Step 12:** Print the new minimum spanning tree with the edges and cost.

**Step 13:** Delete all stored trees with cost < cost of the new minimal spanning.

**Step 14**: End while.

**Step 15**: STOP.

## B. Illustrative example for the execution of the algorithm

Consider the following graph as a test case.

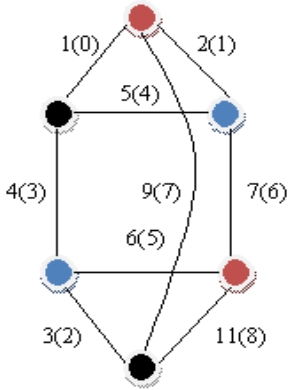

Kruskal's Algorithm may be used to find the minimum spanning tree of the graph and the cost is generated.

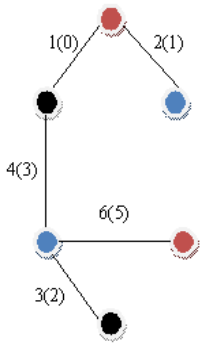

The cost of the minimum spanning tree = 16.

### TABLE I
DWCM FOR MINIMUM SPANNING TREE

|   | 0 | 1 | 2 | 3 | 5 |
|---|---|---|---|---|---|
| **4** | 4 | 3 | 0 | 0 | 0 |
| **6** | 6 | 5 | 0 | 3 | 1 |
| **7** | 8 | 0 | 6 | 5 | 0 |
| **8** | 0 | 0 | 8 | 0 | 5 |

Second minimum spanning tree:

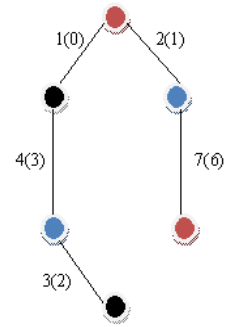

The cost of the second minimum spanning tree = 17.

### TABLE 2
DWCM FOR SECOND MINIMUM SPANNING TREE

|   | 0 | 1 | 2 | 3 | 6 |
|---|---|---|---|---|---|
| **4** | 4 | 3 | 0 | 0 | 0 |
| **5** | 5 | 4 | 0 | 2 | -1 |
| **7** | 8 | 0 | 6 | 5 | 0 |
| **8** | 10 | 9 | 8 | 7 | 4 |

Third minimum spanning tree:

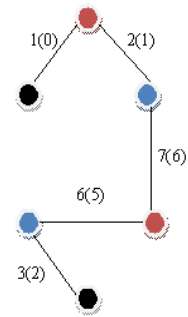

The cost of the third minimum spanning tree = 19.

### TABLE 3
DWCM FOR THIRD MINIMUM SPANNING TREE

|   | 0 | 1 | 2 | 5 | 6 |
|---|---|---|---|---|---|
| **4** | 4 | 3 | 0 | 0 | 0 |
| **3** | 3 | 2 | 0 | -2 | -3 |
| **7** | 0 | 7 | 6 | 3 | 2 |
| **8** | 0 | 0 | 8 | 5 | 0 |

Fourth minimal spanning tree:

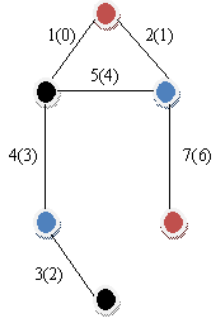

The cost of the fourth minimal spanning tree = 20.

### TABLE 4
DWCM FOR FOURTH MINIMUM SPANNING TREE

|   | 0 | 4 | 2 | 3 | 6 |
|---|---|---|---|---|---|
| **1** | 1 | -3 | 0 | 0 | 0 |
| **5** | 0 | 1 | 0 | 2 | -1 |
| **7** | 8 | 0 | 6 | 5 | 0 |
| **8** | 0 | 6 | 8 | 7 | 4 |

Fifth minimal spanning tree:

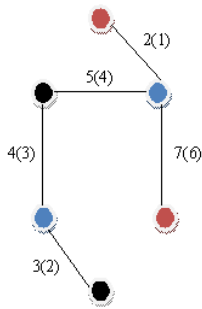

The cost of the fifth minimum spanning tree = 21.

In the same process we can generate the remaining spanning trees.

### C. Proposed theorem:

i. **Theorem1:** *A spanning tree T (of a given weighted connected graph G) generated sequentially in the algorithm OMST is in the increasing order.*

**Proof:-** As the initial MST is generated by Prims's algorithm. It gives the first minimum spanning tree. Then in each iteration the smallest spanning tree is decided from the stored spanning trees (i.e. those have been generated in the previous stage those cost are larger than the smallest spanning tree in this current state) and the spanning trees generating in the current iteration according to the algorithm.

Let T1 be the $k^{th}$ smallest spanning tree in G satisfying the hypothesis of the theorem (i.e. there is no spanning tree T2,(w(T2)<w(T1)) in between the T1 and previous smallest spanning tree).The proof will be completed by showing that if T2 is a shortest spanning tree differ with the T1 in G ,the weight of T1 will also be equal to that of T2.

When iteration K in the enumeration process refers to the iteration in which $1^{st}$, $2^{nd}$,....$K-1^{th}$ iterations are determined .At this iteration a list contains a set of spanning trees with the property that.

i)They are mutually disjoint.

ii)None of the previous stage can generate the spanning tree with lesser weight than itself own weight.

Iii) the union of all spanning trees.

Those have been generated in the previous stage and the K th stage generated spanning trees from the

(K-1)th shortest spanning tree using the OMST algorithm.

ii. **Theorem2:-** Algorithm OMST always generates minimal spanning trees in increasing order**.**

**Proof:-** As we explore all the spanning trees from a single spanning tree using the DWCM and all the positive weighted difference are in count for the next generation spanning trees with higher cost. And all spanning trees generated in the previous stages with higher cost are in consideration for deciding the next level spanning tree. So this algorithm generates all spanning trees in increasing order.

### D. Complexity

Let 'e' be the number of edges and 'n' is the number of vertices and 'N' the number of spanning trees of a given graph G. There are different techniques for generating spanning trees for a graph. Elementary tree transformation is one of them where alternative trees are generated by exchanging one of branch with one chord at a time. Since there are no such order obtained by our algorithm O(N-n) space is need to generate all spanning trees. This limit can't be exceeded due to the generated spanning tree can't cross the value of 'N'. But in most cases only a small fraction of space is needed at any moment to store the intermediate spanning trees.

The time complexity of the algorithm can be calculated as follows where at the initial stage the generation of MST using Prim's is $O(n^2)$. Then in each step it needs to generate the (Fundamental circuit matrix)DWCM[2]. The complexity for that algorithm is $n^k$ where $2<= k<=3$. Then for finding minimum positive value for each row takes time $O(n-1)$ and the exchange is done by constant time. So the total time complexity is $O(n^2 + N \cdot n^{k+n-1})$.

As the value of N is exponential its time complexity is exponential in theoretically but in practical cases its number never crosses the fraction of the value of 'N'.

### E. Application:

This proposed algorithm is mainly in the class of MST problem with additional constraints. That is degree constrained MST, Hop constrained MST, weight constrained MST etc. These are all NP-Complete nature.[g/j]

We proposed an algorithm which can be applied in the mobile computing when it finds an congestion in the MST and it needs Immediate next MST. This algorithm gives suitable result. This algorithm can be applied in the various application of routing algorithms.

### V. CONCLUSION

The area discussed here is known to us. The minimal spanning tree represents the minimal path between the nodes of the graph. It may possible some times in real life that minimal path can't be reached due to some circumstances, in that case the next minimal spanning tree is useful. So we hope that this contribution will benefits some areas of real life problem. At a first look this algorithm may seems complex but it is as simple as it can be performed in paper and pencil.